# Personal vs. Know-How Contacts: Which Matter More in Wiki Elections?

Yousra Asim, Muaz A. Niazi*, Basit Raza, Ahmad Kamran Malik

Department of Computer Science, COMSATS Institute of Information Technology, Islamabad, Pakistan

engryousraasim@gmail.com, muaz.niazi@gmail.com, basit.raza@comsats.edu.pk, ahmad.kamran@comsats.edu.pk

*Corresponding Author: Muaz A. Niazi

**Abstract**

The use of social media affects the real world as well. As an example, an election candidate's social media coverage can affect the chances of success. As an example, literature has identified that the same can be predicted by the online public discussions of online social network users about a candidate, besides the use of Twitter by a candidate himself. The contacts of a candidate in terms of number of online fans on Facebook, and number of followers on Twitter are found playing a vital role behind his success. Among many other forms of social media, Wikipedia is a very widely used channel. It is a freely available platform for writers all over the world where people can write and edit articles. Some of its users have more technical access for administration. Such users are selected by carrying out the nominating process in online public elections. The online election data are quite interesting in terms of being an emergent outcome of a large-scale self-organized opinion formation process. However, due to dynamical, and non-linear interactions and presence of mutual dependencies between election participants, the analysis of this data can be cumbersome using traditional statistical techniques. Also, the data of these elections are large and it is difficult to find the exact patterns in this data. For this purpose, social network analysis seems a better alternative for the identification of local and global patterns, locating influential nodes and the contacts behind their influence and examining social network dynamics. As a consequence, this study has to rely on specific social network measures to investigate the interactions between election participants and the importance of their contacts. It investigates whether personal contacts matter more than know-how contacts in wiki election nominations and voting participation by using standard tools such as Pajek and Gephi. It further evaluates the significance of a person's contacts in online wiki elections through a number of different graph-based influence identification methods. Additionally, the basic characteristics and cohesive groups in the wiki vote network are explored. This work contributes by discovering the significance of personal contacts over know-how contacts of a person in online elections. It is found that personal contacts, i.e. immediate neighbors (degree centrality) and neighborhood (k-neighbors) of a person have a positive effect on a person's nomination as an administrator and also contribute to the active participation of voters in voting. Moreover, know-how contacts, analyzed by means of measures such as betweenness and closeness centralities, have a relatively insignificant effect on the selection of a person. However, know-how contacts in terms of betweenness centrality for passing information in the network can positively contribute only to the voting process. These contacts also measured in terms of influence domain and PageRank can play a vital role in the selection of an admin. Additionally, such contacts in terms of reachability and brokerage roles have a positive association with the voting process.

## Introduction

The importance of Online Social networks (OSNs) cannot be neglected as it facilitate individuals to communicate and share their opinions and ideas with each other. People are constantly involved in using different available OSNs such as Twitter, Facebook, etc. Along with several other sharing activities, users also indulge in election related happenings. Elections can be of different kinds spanning from local to global e.g. General election (Burnap et al. 2016), Senate election (Smith and Gustafson 2017), Parliamentary election (Smith and Gustafson 2017), and Wikipedia election (Jankowski-Lorek et al. 2013). The election related activities performed by voters (Bode et al. 2014) and candidates (Brady et al. 2017) can predict forthcoming results. These predictions of election results can be derived by using the data of users of OSNs in many ways. For example, user voting intentions can be predicted by the analysis of a user's online comments and text about a particular candidate; user opinions can be collected online about expected election results; the data of pre-election polls can also be helpful in this context (Harald et al. 2013)

It is interesting to note how the use of OSNs can greatly influence the success rate of a candidate in election. Researchers have suggested to explore the impact of the structure of OSNs for electoral outcomes. Some studies have considered the use of underlying social network structure of online elections to investigate the electoral outcomes. For example, the literature provides evidence that the voters who are more connected to each other can provide better prediction of group voting behavior in elections (Conitzer 2012). It indicates the influence of contacts of a candidate on his voting behavior. Personal and Know-how contacts of a candidate are important and can be analyzed by using different centralities such as closeness centrality, degree centrality, betwenness centrality etc.

Another popular social network platform is Wikipedia, which facilitates worldwide writers to work together, who have a shared goal of providing knowledge to the community (Sheth and Kapanipathi 2016). This work is related to Wikipedia platform only. A few individuals have greater technical access and can maintain different features and quality of content on Wikipedia (Lee et al. 2012). These people are called administrators. Different factors involved in Wikipedia online elections (conducted for administrator selection) are identified by exploring contacts based social network structure of wiki vote network (Lee et al. 2012). Different predictors that can lead a candidate to success are highlighted. It is seen that most studies are performed by using statistical analysis in online settings e.g. (Kordzadeh and Kreider 2016). A few SNA-based studies in this perspective, have extracted a set of features from the social network of voters e.g. (Oppong-Tawiah et al. 2016). Further, these features are used for prediction of a successful candidate by using logistic regression classifier. The data used in these studies consists of positive, negative and neutral votes of voters. However, the work that focuses on the communication pattern (social links) between participants during online elections is rarely available. The contacts of a person which matter more in Wikipedia elections are not investigated. The significance of personal contacts and know-how contacts of a person by using SNA algorithms is also unexplored.

The scope of this work is limited to investigating the patterns of the RfA process. Our main objective is to examine the importance of personal and know-how contacts of participants in wiki elections. For this purpose, an underlying social network of wiki elections is considered to investigate the importance of social contacts between participants. A number of different SNA algorithms are used for this purpose. Degree centrality and k-neighbor algorithm are used to find personal contacts of each participant of the wiki elections. The algorithms used to find know-how contacts of a person are closeness centrality (Sabidussi 1966), betweenness centrality (Freeman 1977), k-core (Seidman 1983), PageRank (Page et al. 1998) , and brokerage role (Burt 1992). The rest of the paper is organized as follows: The section named “Methods” represents the description of the dataset and a brief description of used algorithms. Detailed analysis of dataset is performed in “Results”. “Discussion” section includes the details about findings and previous relevant studies. At the end, the “Conclusion” section concludes the paper.

## Methods

In this paper, the following steps are performed for the solution. First, the literature review is conducted and a dataset of the Wikipedia vote network[1] is selected from publicly available data-sets of Stanford University. Afterwards, data analysis is performed by means of two standard SNA tools named Pajek and Gephi. Different social network measures are employed to examine the patterns of the network under investigation. The details of the selected dataset are as follows:

### Dataset Description

In this paper, a Wikipedia voting dataset is used that was gathered from 3 to 31 Jan. 2008.It has also been used in (Leskovec et al. 2010a) and (Leskovec et al. 2010b). Both of these works have focused on the positive and negative links between participants of wiki elections. However, instead of using signed information of the wiki vote network, this work is only focused to positive links information. Data about 2794 online wiki elections were collected out of which 1235 elections were successful. This dataset consists of the information regarding votes given by previous admins as well as wiki users. There are 7115 voters (nodes) and 103,663 votes (arcs) in the network. It is a directed graph where A->B means that node A votes on node B; A is a voter and B is the nomination. Also, B can give a vote for A without restriction.

[1] http://snap.Stanford.edu/data/wiki-Vote.html

**Formal description of important algorithms used**

A number of different relevant well-known graph-based measures are selected to achieve the research objective of this study and used for data analysis, which are described as follows.

- Degree centrality of a node determines the number of ties of a node with its immediate neighbors. The more the number of neighboring nodes are, the more important the nodes are. Every neighboring node is considered as one “centrality point” (Batool and Niazi 2014). Though, degree of importance of neighboring nodes may vary and important neighbors contribute towards the importance of a node (Khan and Niazi 2017). Degree centrality can help to examine the personal contacts of participants because it focuses on direct connections of a node with its neighbors (one hop neighbors). The high degree centrality of a node can indicate its large neighborhood of direct contacts.
- A measure of centrality that calculates the sum of the length of the shortest paths between a node and all other nodes in the network is known as Closeness centrality (Sabidussi 1966). It represents the extent to which an individual node is close to other nodes in the network. This measure highlights the individuals who are best placed in such a way that they can rapidly influence the entire network. It can be used to analyze the know-how contacts of participants because the closeness centrality for a single node is calculated by considering all other nodes of the network. It does not include only immediate neighbors of a node.
- Betweenness centrality finds the importance of a node in any communication in the network. It finds the extent to which a node is involved in the geodesics of the pairs of the other nodes in the network (Batool and Niazi 2014). High betwenness of a node can depict its authority over other nodes in the network or mentions its hold in collaboration between other nodes[2]. It can show the acting bridges in the network communication. As this measure includes the chain of nodes that are found between the communication of any two nodes, it can be used to represent the know-how contacts of a candidate.
- K-core represents a maximal subnetwork in which each node is connected to at least k other nodes within that subnetwork (Seidman 1983). A node can be associated to multiple cores. Moreover, nodes of one core can belong to a number of components, which indicates that a k-core can be a disconnected subnetwork (Douglas A 2016). The importance of contacts of a candidate can be explored by examining the nodes present in the highest k-core of the network. Because each node in this subnetwork has the highest probability to be connected with k others in that subnetwork.

[2] https://cambridge-intelligence.com/keylines-faqs-social-network-analysis/

- Centralization is used to depict the entire network while centrality is related to a single node. A highly centralized network indicates a clear margin between the center and the border of the network (De Nooy et al. 2011). It indicates the extent to which a graph is tightly organized around some most central points. A point will be globally central if it has large neighborhood of contacts with respect to whole network. This measure can facilitate to find the solidarity of the graph by analyzing the prominance of indirect contacts. Degree centralization determines the extent to which the organization of nodes in the network is around some central points. Betweenness centralization finds that to what extent the nodes of a network are important in information flow in that network. Closeness centralization of a network describes the organization of nodes in strongly connected groups.
- K-neighbor determines the distance of a selected node from all other nodes in the network. Besides, some closest or some distant nodes of the selected nodes can be found as well (De Nooy et al. 2011). If the shortest path from node A to node B has length k then node B is its k-neighbor. The larger neighborhood of closest other nodes (dense neighborhood) or smaller neighborhood of distant nodes with respect to a particular node can help to investigate the importance of its know-how contacts.
- Louvain algorithm divides a network into high-quality partitions using modularity measure (Blondel et al. 2008). Modularity is a well-known metric that compares the number of connections inside a partition and between different resulting partitions of a network. High quality partitions exist if there are more connection found within partitions than between them (Asim et al. 2017). As it is discussed earlier, that non-linearity in the number of connections exist in the social network, community detection can help to study the local structure of the network. The more number of connections within a community and the size of that community (possible contacts) can be positively associated with success of a candidate.
- PageRank calculates user's centrality scores based on their connectivity in the weighted activity graph where a weighted activity graph is derived by using user's communication activity and the strength of users' connections (Heidemann et al. 2010). This measure can be used to uncover those nodes who are influential beyond their immediate connections. The connections of a node with such vital nodes can positively contribute towards its influence in the network. So, this measure can provide insight into the significance of know-how contacts of a candidate in electoral outcomes.
- Structural hole is determined by focusing on the individual node (ego) instead of complete network. The node under consideration is called ego. If there are three nodes A, B, and C in the network, and A is connected to B and B is connected to C, then the absence of tie between A and C represents structural hole. It means that A is dependant on B to communicate with C (De Nooy et al. 2011). Here, B can play a role of agent or broker between A and C due to the presence of structural hole between them. Brokerage roles represent the position of a broker in an incomplete directed triad. Broker takes advantage from its postion between the nodes who are not directly connected to each other. Triad represents a subnetwork of three directly connected nodes. If two members of a group use mediator v from outside, then v is called itinerant. If a member regulates flow of information from her/his group to another, then, s/he is representative. If a node regulated flow of information to her/his group then, s/he is called gatekeeper. A person who mediated between members of diff erent groups being not the part of those groups is called liaison. Lastly, if a mediator is also the member of the same group and part of communication as well then s/he is the coordinator (Gould and Fernandez 1989). Generally speaking, the persons playing brokerage roles can be referred as transmitting actors. Such actors are the only way of the communication between two parties. They can offer pivotal role in the network and can be explored in the context of online elections in determining their worth for a candidate (Täube 2004).

By using the aforementioned social network analysis measures, we have been able to explore the importance of contacts of a nominee in Wikipedia online elction happenings.

## Results

This section represents the results found during analysis of the dataset and discussion on these findings is also provided.

### General characteristics of the Network

The following wiki-vote network as shown in Fig. 1 has some general characteristics which are described below.

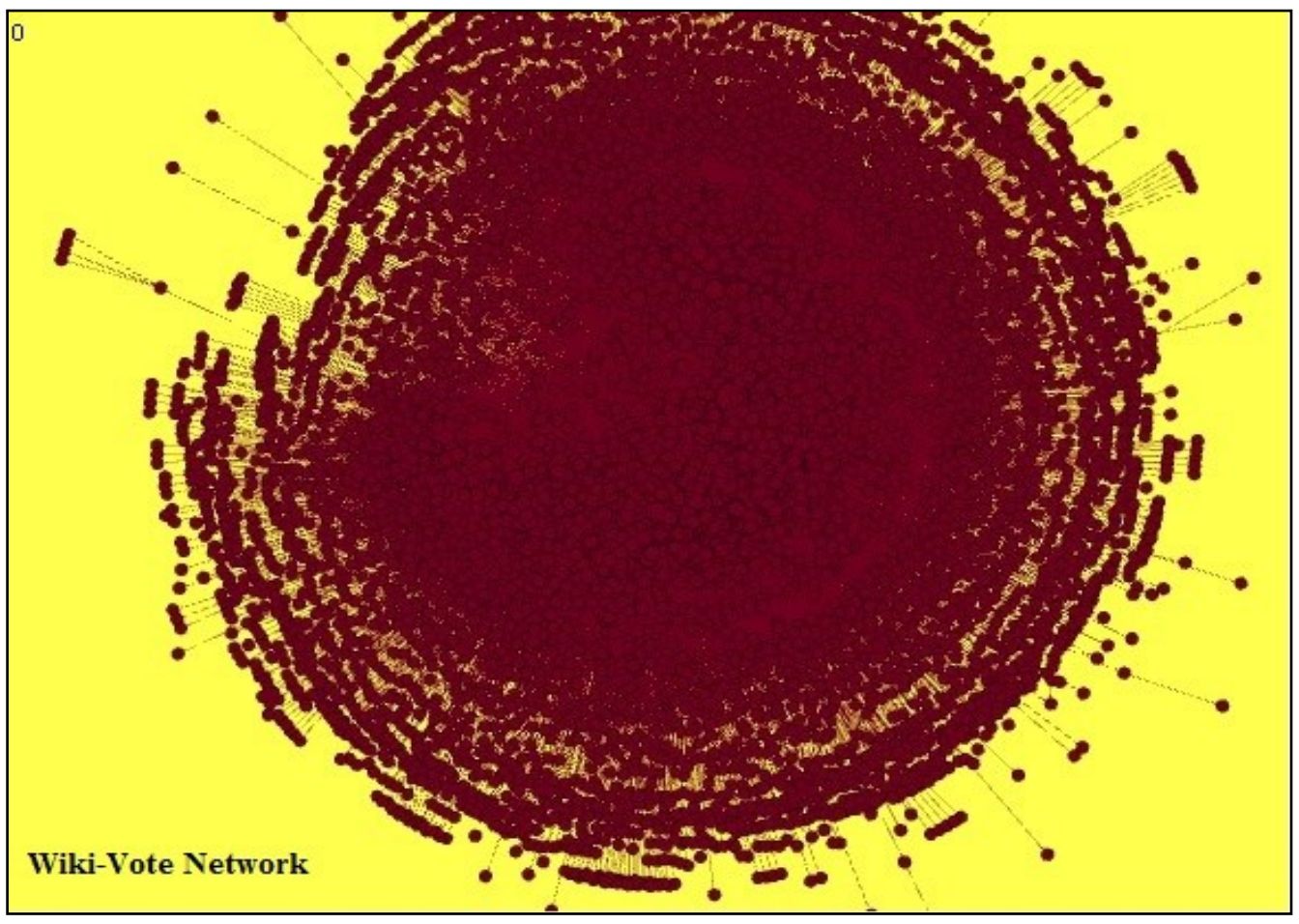

Fig. 1. Wiki-vote Network

First of all, it is directed network as arc will show who selected whom as admin. It is a single mode network because all nodes are of the same type i.e., persons. It is a simple directed graph because it has no multiple arcs (a person is selecting one person for only once). Moreover, the network has 7115 nodes (nominations as well as nominators) and 103689 arcs between all nodes. An arc A->B represents that A is a voter and B is the person whom the voter wants to select as an administrator. In this network, there are no loops found which means that a person cannot select himself as an admin. The average degree of the network is 29.146 which shows that nodes in the network have nominated more than 29 other persons to be an admin. The density of the network is very small that is 0.00204854 which shows that only 0.2048 percent of all possible arcs are present. It means that this network is not dense. The reason behind this can be the large size of the network, which can reduce the possibility of the presence of maximum arcs because voters are not selecting every other person as admin. Diameter shows the longest shortest path. It is clear in Fig. 2 that diameter of the network is 10 which is the average length of shortest path present in the network. It also shows the path between both farthest away persons of the network (from V624 to V3592). It indicates that there is the longest chain of ten know-how contacts between two persons at maximum in the network.

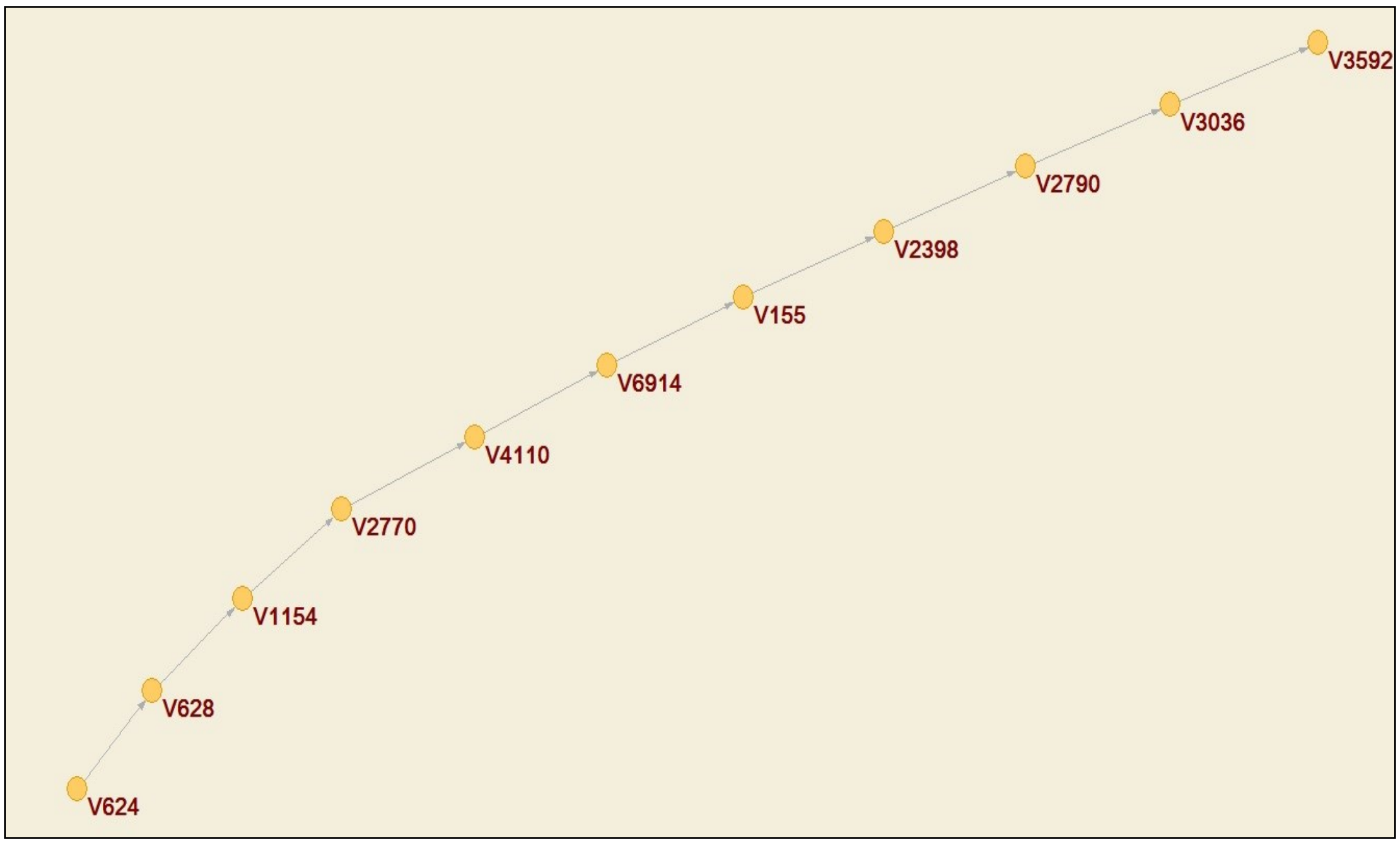

Fig. 2. Network Diameter

**Centrality effects on admin-ship and on active voter**

OSN is a collection of nodes where the degree of importance of each node can vary in the network (Yang and Xie 2016). Some nodes can have more influence or importance in the network than others. Such nodes are called influential nodes. In the context of wiki-elections, as voters have the power of nominating admin by giving their vote to the person of their choice, the most active participants in voting can be considered as influential. In the nomination process, some candidates will be highly wanted as admin than others. Such candidates are also vital in elections because any one of them will be selected as an admin among thousands of people. To find out such influential persons, degree centrality algorithm, betweenness centrality algorithm, and closeness centrality algorithm is used.

Input degree of nodes varies from 0 to 457 in the wiki-vote network. There are 4734 persons who have not been voted, not for even once (i.e. having 0 in-degree. Furthermore, a nominated person can have many votes from voters (in-degree). If a person has a number of votes, then it means that he has a high chance of being nominated as admin. It shows the importance of personal contacts of a person for being selected by voters. Results show that node v4037 is the most wanted administrator (i.e. has 457 in-degree). It indicates that personal contacts of a person contribute to her/his importance in elections. Fig. 3 shows the nodes w.r.t. their in-degree centralities where the size of a node represents its high degree centrality.

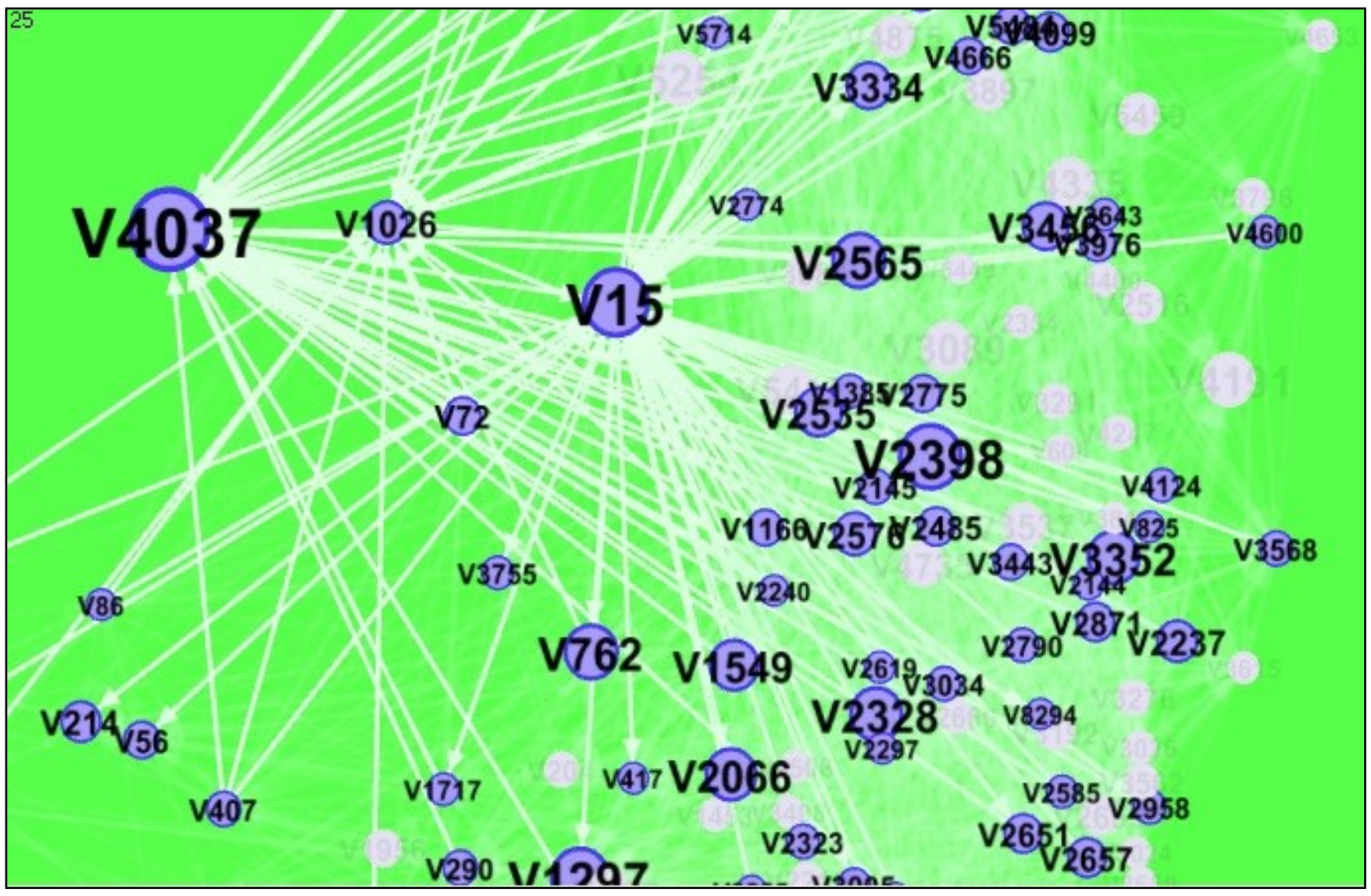

Fig. 3. In-degree Centralities of Nodes

Another influential found in elections is node v2565, who is the most participative voter (i.e. with a maximum out-degree) as mentioned in Fig. 4. It is the most active node in the network in nominating admins. This indicates that personal contacts of a person to/from other nodes affect online elections and positively associated with administrator selection and voter participation.

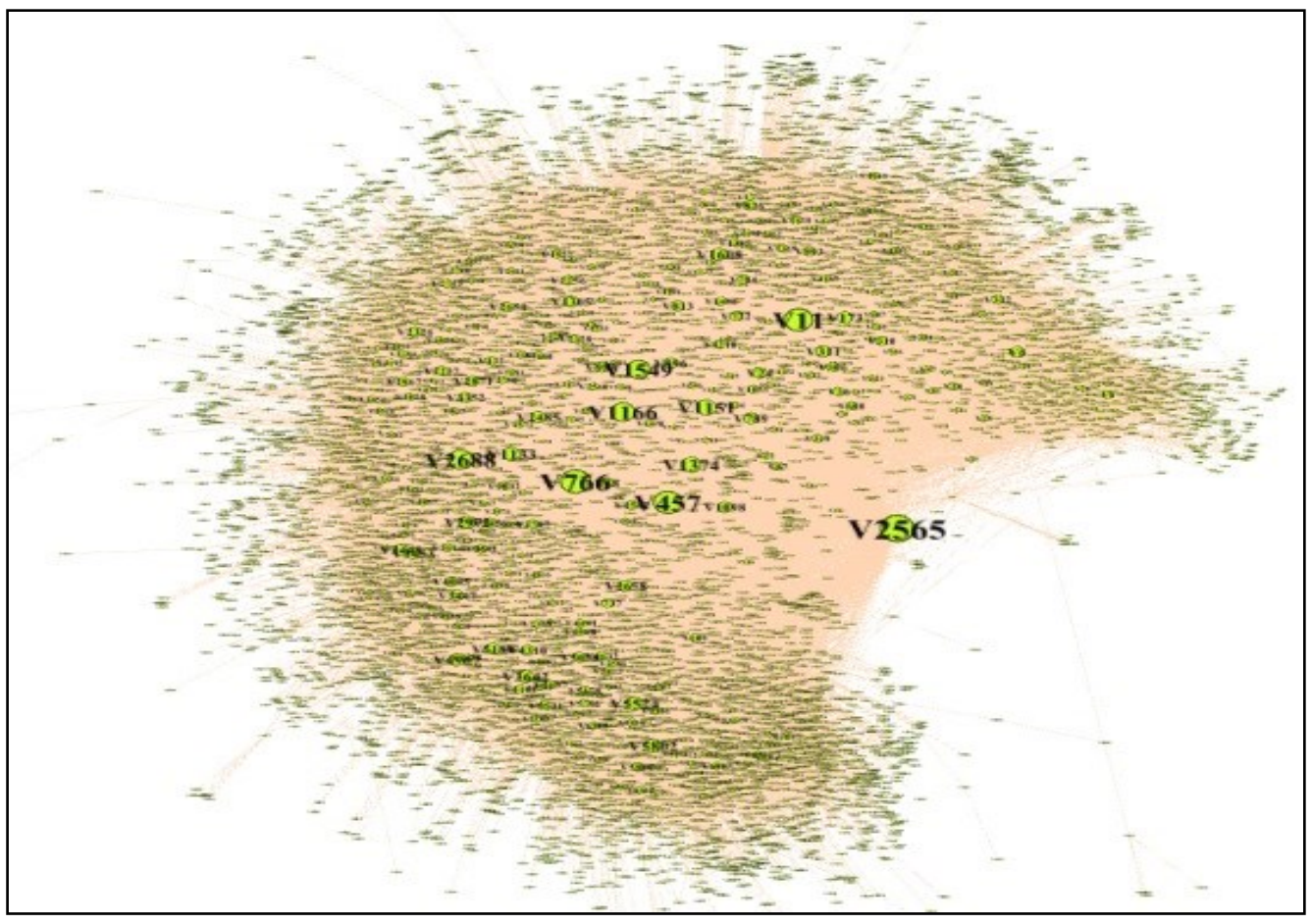

Fig. 4. Out-degree Centralities of Nodes

Additionally, the overall degree centrality results show that people in the wiki-vote network are more active in giving votes to others than being nominated by others. The degrees of nodes also show that 4734 nodes are not nominated by others and 1005 nodes are not participating in the voting process. There are 65.35% people who are not nominated by others and 14.12% people who are not giving their votes to anybody. There is a 51.23% increase in voting behavior in elections than being nominated. The Fig. 5 shows the distribution of out-degrees of nodes. Here, the x-axis represents the out-degree (participation) of a node in the election and the y-axis represents the number of nodes. It is clear from x-axis that a few nodes are having a minimum degree and lie near the origin on the x-axis. Further, most of the nodes lie on the greater values of x-axis having greater out-degrees.

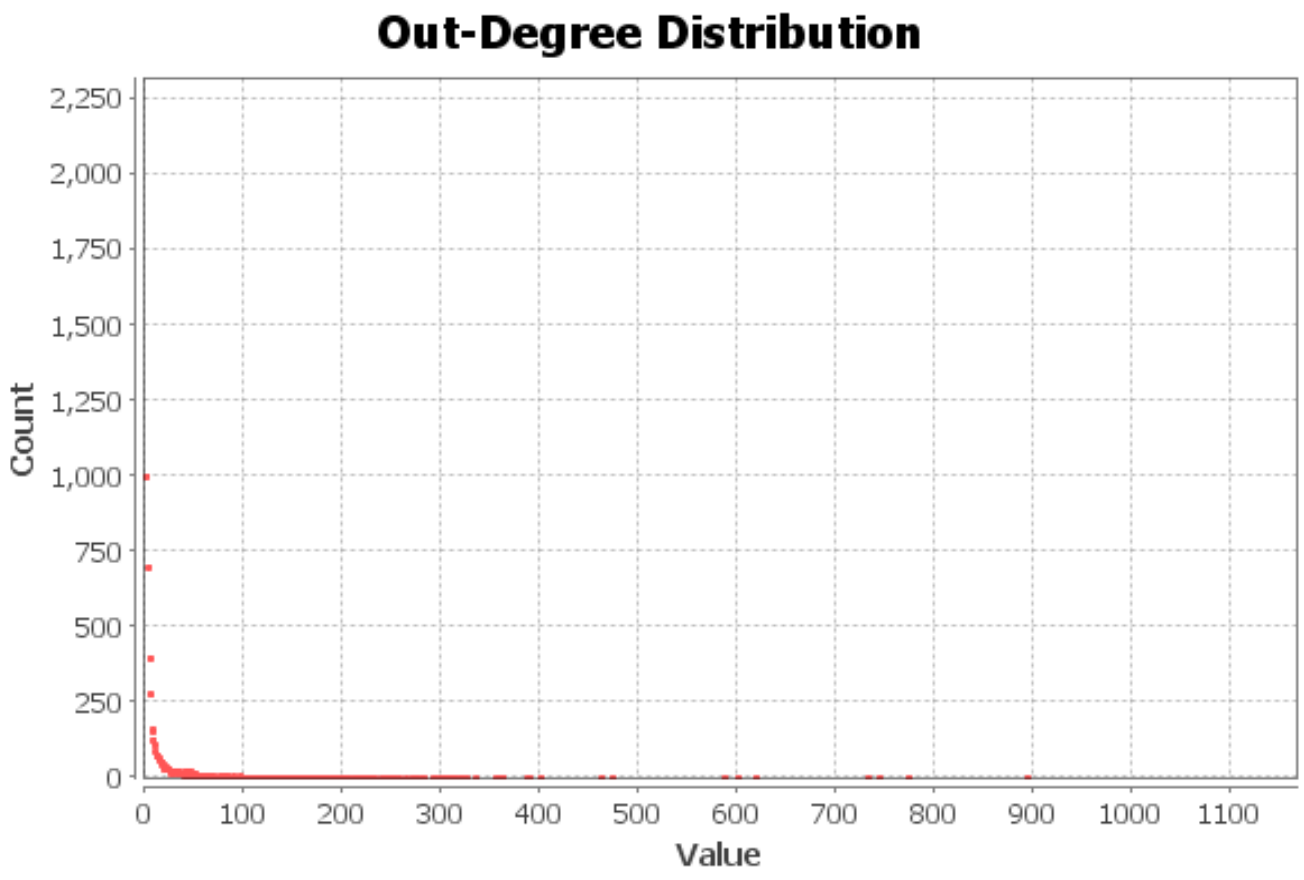

Fig. 5. Participation of Nodes in Election

Betweenness centrality is also an important measure to find the intermediate nodes from which data flow from one person to another (Kitsak et al. 2010)(Kiss and Bichler 2008). If more data pass through a node, then that node will remain aware of other person's choices. Such a node can most frequently control information flow in the network. This measure can find the influence of the know-how contacts in wiki elections. The node v2565 is found with the maximum betweenness (i.e. 0.0612). There are 5740 nodes in the network that are not playing the intermediate role in the data flow in the network. Fig. 6 shows the sizes of nodes w.r.t their betweenness values where v2565 (the most influential voter) has the largest size. The larger the betweenness value of a node; larger the size of a node. It highlights that due to the high probability of node v2565 presence in the information flow of the network, s/he has more capability of giving the vote to more persons than other nodes. This way, know-how contacts are influencing positively to the voting process of wiki elections.

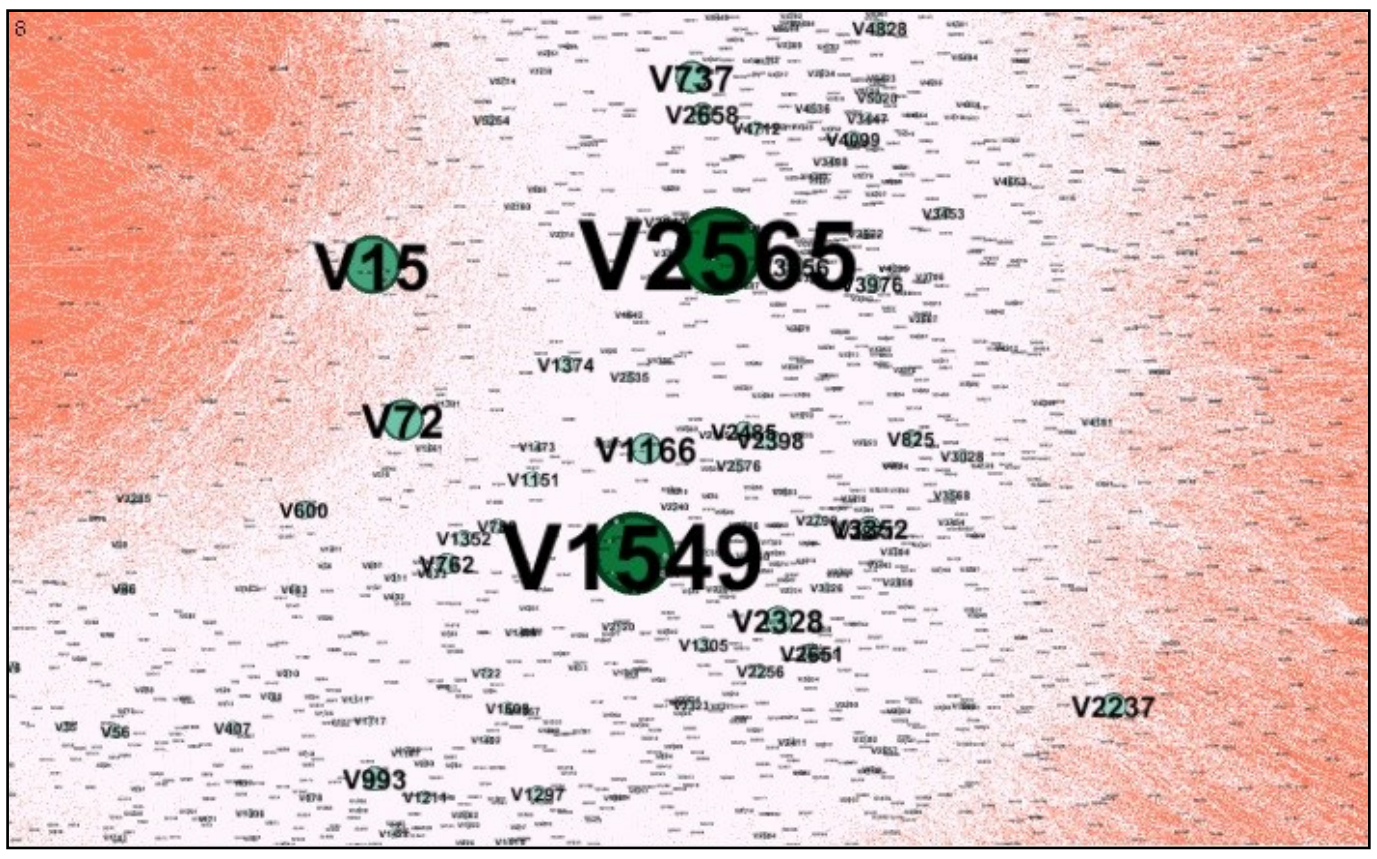

Fig. 6. Betweenness Centralities of Nodes

Closeness centrality shows the distance of a node from all other nodes (Sabidussi 1966)(Kiss and Bichler 2008). If a node is close to all other nodes, then information will reach it very quickly. In this network, v628 has maximum closeness centrality which means that s/he can get the result of election more quickly than others when it will be announced. This person can vote any person more quickly than other voters and s/he has also the chance to be selected as admin more rapidly. The size of v628 is bigger than other nodes in Fig. 7 which indicates its closeness to other nodes in the network. As compared to the results of personal contacts, the findings from closeness centrality specify that in online elections, it does not matter how much nodes are close to each other. The importance of know-how contacts with respect to closeness centrality of a node has no effect on voter participation and admin selection processes.

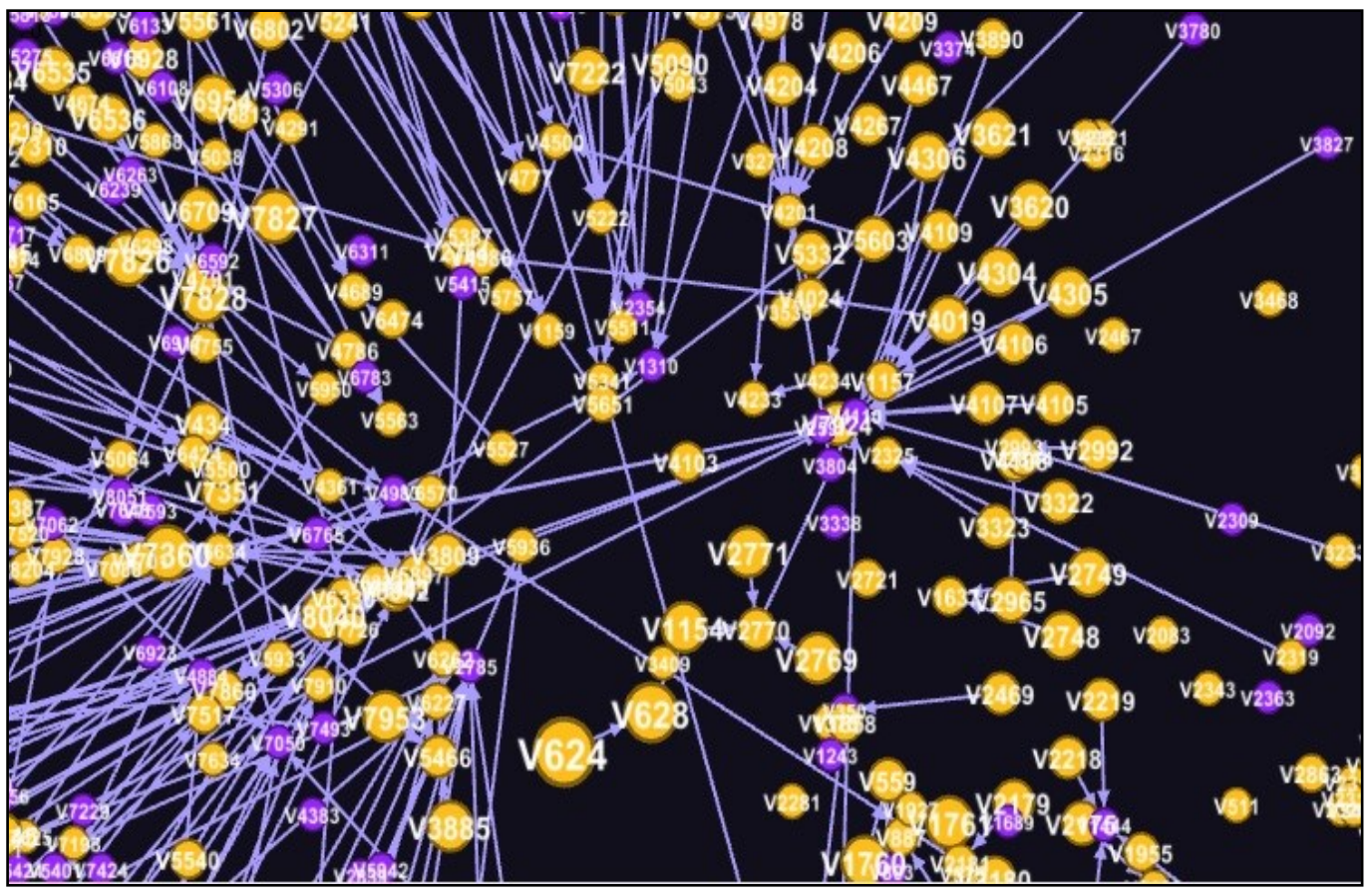

Fig. 7. Closeness centralities of Nodes

**Most cohesive/most connected group of persons in the Network**

A social network can be partitioned into subnetworks to investigate the solid groups in which individuals have more and frequent interactions with each other. The reason behind such grouping can be the similarity of interests and beliefs. The cohesive group is also known as a component. To find such dense groups in wiki-network, the k-core algorithm is used. It says that influential nodes are present in the highest-core of the network (Kitsak et al. 2010). A group of 336 election participants is found the most cohesive in the wiki network, which can be seen in Fig. 8. Furthermore, it is found that this group is 53 k-core. It means that the degree of each node in this group is 53. Every person of this group is connected to other 53 persons. For example, it can be seen in Fig. 9 that V407 is connected to 53 highlighted nodes.

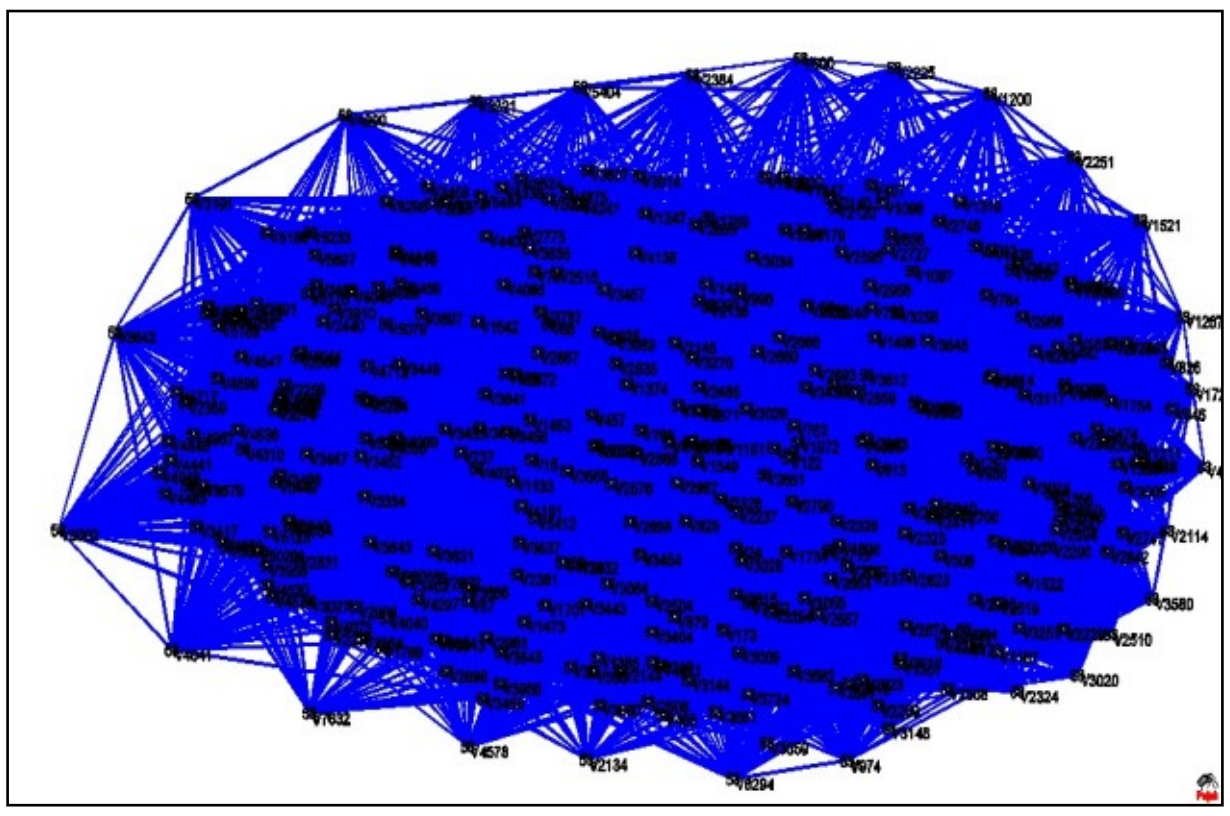

Fig. 8. Most Cohesive group in Wiki-vote Network

The most important point to discuss here is that the most active voter and most wanted admin both are also found in this group. It highlights the significance of contacts of a person in wiki elections. The persons who are more connected to others can have more chance to get nominated as an admin. Also, the voters who have a high number of contacts can actively participate in the online election voting process.

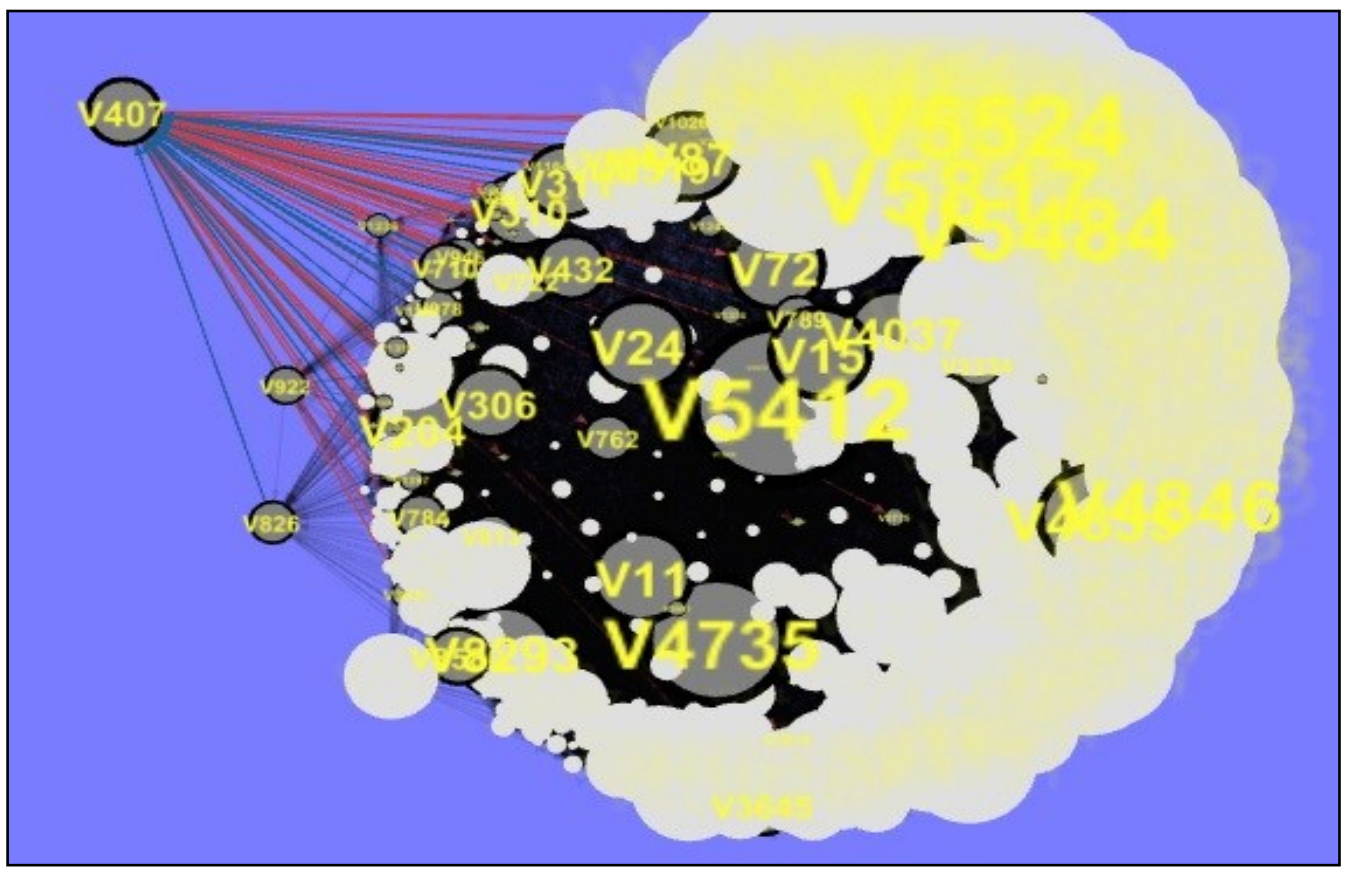

Fig. 9. K-core in the Network

**Network Centralization**

Centralization of a network shows that to which extent some nodes (participants) of a network are of more importance than others. It tells whether the nodes in the network are organized around its most central points. If there exist some central points in the network, then it is called centralized network otherwise decentralized. The maximum value for centralization of a network is 1. The wiki network is found to be non-central (i.e. degree centralization is 0.1457 and betweenness centralization is 0.060). It means that all the nodes are almost at the same level of importance. It is found that centralization with respect to degree and betweenness of nodes has no eff ect on wiki elections. Likewise, closeness centralization cannot be found in this network because it is not strongly connected network. It means that the nodes of the network are not much closer to make groups that are strongly connected.

**k-neighbors of famous nodes**

K-neighbor highlights the distances of all nodes from the selected node and vice-versa. In wiki-vote network, two persons v4037 (the most wanted person as admin) and v2565 (most active person in the voting) are the most important so we are interested in the analysis of their neighborhood on k- hops. The influence domain (number of steps required to reach a particular node) of most wanted admin and reachability (number of steps required by a node to reach other nodes) of most active voter are determined.

The influence domain of v4037 varies from 0 to 7. There are 457 persons who are the closest neighbors of v4037 (at the distance of one hop only). There is just one node that is farthest away from v4037 (at 7-hops). There are total 5158 persons that are found in the influence domain of v4037 which is almost 72% persons of total voters. It means that know-how contacts (dense neighborhood) of a most wanted admin can be the reason behind his /her selection and more voters in her/his favor. Further, he has a wide influence range in her/his neighborhood, even they are far away (at $7^{th}$ hop). Also, most of her/his neighbors are closer to him. Now, the reachability of the most active voter is under consideration which varies from 0 to 4. There are 893 nodes at one hop distance from node v2565 and only eight nodes are at a maximum distance (at 4 hop distance). There are 32% persons of the total population in the network that are reachable by the most prominent voter. Results show that neighbors of this person are organized in a few number of steps. The reason behind her/his active participation, in this case, might be that s/he knows more persons very closely in the network.

**Community detection in the network**

A network can be divided into a number of communities/groups. Community property of a network indicates that there is more number of connections in the communities as compared to the number of connections between them. Researchers have proposed many community detection algorithms. Twenty-nine number of communities are found in the wiki-vote network by using Louvain algorithm (Blondel et al. 2008). Louvain returns the communities where each node in the network is a part of one community only (non-overlapping communities exists). One interesting result found in this analysis is that the community where most wanted admin v4037 is involved has a maximum number of nodes 1944 out of 7115 (almost 27.32% of the whole network). It highlights the fact that a number of connections and number of voters/contacts both are important to get adminship in online elections. These partitions of the network can be seen in Fig. 10 as shown below. Different colors of nodes represent different communities. The nodes having the same color indicate a single community.

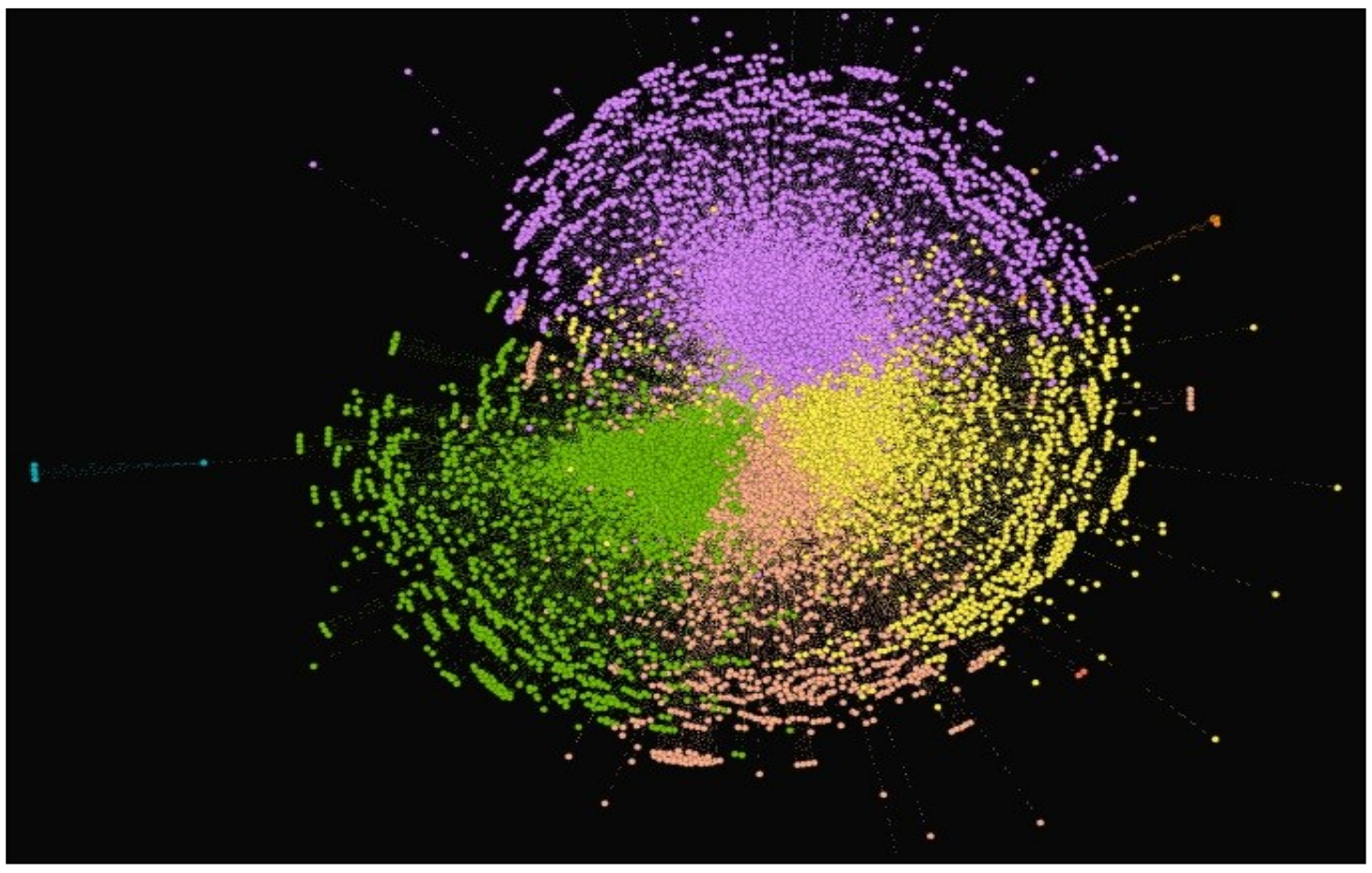

Fig. 10. Community Detection in the Network

**Page Rank**

Page Rank finds the importance of a node by using number of contacts of a node and the importance of those contacts (Heidemann et al. 2010). A node is considered important if it is connected to other vital nodes in a network whether they are small in number. The following Fig. 11 shows that node v4037 is on the highest rank in the network. It demonstrates that this person is connected to a number of important nodes in the network, which can be the fact behind its popularity among voters as the most suitable candidate for adminship.

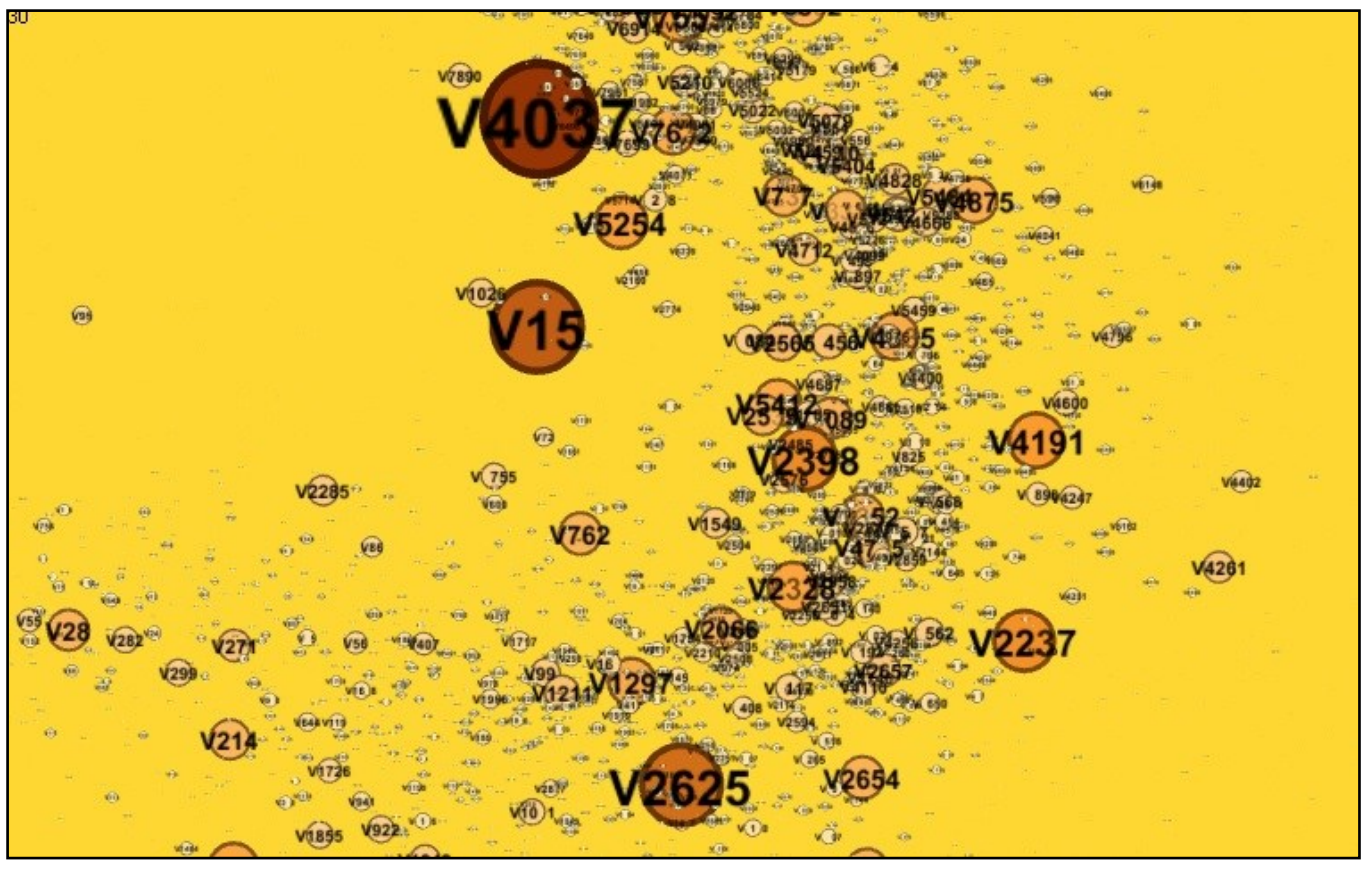

Fig. 11. Ranking of nodes in the Network

**Brokerage roles in network**

A broker is a person who can act as an agent in the communication between different groups. A person can play five different brokerage roles in social networks (De Nooy et al. 2011)(Gould and Fernandez 1989) which have been discussed earlier. To find brokerage roles in the network, the partition of nodes into groups is mandatory. That is why first strong components in the network are found and afterward brokerage roles are determined. It is seen that the largest strong component in the network is only one having 1300 nodes. Table 1 provides information about brokerage roles of wiki network. Pajek calculates the five brokerage roles for each node in the network. The results show that the most active voter (v2565) is playing the most number of roles in the network except itinerant. As we have seen earlier, that this node has the highest out-degree (connected to 893 others) while an average degree of a node in the network is 29.14 only. The degree of v2565 is much higher as compared to other nodes. The high number of contacts of node v2565 and more opportunity for being an agent both can be the reasons behind her/his high participation in elections. Because, being an agent, s/he can be a part of many communications between members of different components. This way, s/he knows a number of persons and can give votes to more people. It indicates that know-how contacts in terms of brokerage roles can affect voting in online elections.

TABLE I. BROKERAGE ROLES IN THE NETWORK

| | **Representative Role** | **Gate Keeper Role** | **Liaison Role** | **Coordinator Role** |
|---|---|---|---|---|
| **Total Roles in** | 665007 | 1169062 | 550529 | 1334802 |
| **the Network** | | | | |
| **Persons without Roles** | 5994 | 5832 | 5928 | 5818 |
| **Persons with Roles** | 1121 | 1283 | 1187 | 1294 |
| **Person playing the most number of roles** | V2565 (38450 roles) | V2565 (73900 roles) | V2565 (37508 roles) | V2565 (68624 roles) |

## Discussion

The success of a contestant in online elections has been an area of interest for the research community. Due to vivid participation of users in OSNs, their data on these platforms has been investigated in the perspective of election consequences. The candidates of election can target virtual communities to gain attention of the audience. Social media has a direct link with traditional media. A review of the relevant literature provides a strong evidence that the candidates who are prominent in traditional media also receive more number of mentions on social media like Twitter (Hong and Nadler 2012). Because of this impact, the candidates are advised to invest early and heavily in social media campaigns in case of high competition during elections for effective results (Cameron et al. 2016). The success of Obama in the 2008 elections could have been successfully predicted by noticing his widespread appeal amongst American youngsters on Facebook and MySpace (Dalton 2009). Likewise, the claims of US President Donald Trump enforces this idea of the power of social media when he said that Twitter powered him to directly target voters for being successful in the 2016 US election (Brady et al. 2017). Besides, Twitter data has been used for the reasonable prediction of electoral outcome before UK 2015 general election (Burnap et al. 2016). The use of Twitter in elections by a candidate is found directly proportional to his electoral support, i.e. the number of votes received by that candidate (Kruikemeier 2014). It is has already been noticed that the candidate who is receiving more attention in public discussions on Twitter can receive more votes in elections (DiGrazia et al. 2013). The candidates of election can have a number of different contacts on these platforms. These contacts can be friends, followers, friend of a friend etc.

This study has investigated the importance of contacts of a person (whether they are direct or indirect) in online wiki elections by means of a number of different graph-based measures. We have used the methods of influential node identification based on a local measure named degree centrality and k-neighbors to find the direct (personal) contacts of a person. Likewise, different global measures are used to find the indirect (know-how) contacts of a person named betweenness centrality, k-core, page rank, and brokerage roles. The results of the election are examined in terms of the success of a candidate (admin selection) and involvement of a candidate (participation in voting). It is found that direct (personal) contacts of a person have significance in both cases. Apart from these results, research community has also shown that contacts and actions of those contacts of an individual, influence her/his participation and decision-making process respectively in voting on Wikipedia (Lee et al. 2012). Furthermore, our results also seems consistant with the findings of (Lin 2017) and (Spierings and Jacobs 2014) about the importance of contacts of a candidate in electoral outcome on other OSNs e.g. Facebook and Twitter. It is explored that importance of contacts of a candidate in terms of number of fans of a candidate on the Facebook in municipal election carried out in Taiwan. It is found that number of online fans of a candidate is positively associated to the election outcomes (Lin 2017). It has been examined, that the number of followers of a candidate positively contributes towards preferential voting against him, provided that candidate is actively participating in OSNs (Spierings and Jacobs 2014).

A finding of this work shows that the neighborhood of a person is also a vital factor behind his selection as an admin and his active participation in elections. Also, the results of the most cohesive group in the network determined by k-core, highlight the presence of most influential nodes of wiki election. It indicates that the position of a person in the network (with respect to contacts in the k-core) makes him prominent in his selection and participation during elections. The literature has evidence that the actions of the contacts of a voter and the support of important voters can affect the decision of voters in voting (Cabunducan et al. 2011). Besides, the study (Oppong-Tawiah et al. 2016) has investigated the effect of both *structural capital* (degree and eigenvector centrality) and *social connectedness* (sentiment score and sentiment similarity index) of a candidate on an individual and community level. It is found that eigenvector centrality of a person from voters can make him successful to achieve adminship. It indicates that the important contacts of a person are positively associated to his success and it is worth mentioned here, that in highest k-core, all nodes are connected to other k nodes (maximally connected in that sub network). In this way, the support of the most important voters of the network to a person leads him to get selected for adminship.

On the other hand, our results show that indirect (know-how) contacts of a person contributes to his influence in some way in an online election. Results indicate that if a person is important in the communication between other persons of the network, then s/he is important in the voting process. Burt has elaborated that the access of a person in a network can be determined by the network around him (Burt et al. 2013). He considered a person as “ego” which is surrounded by a number of contacts, who are not directly connected to each other. In this work, we find that the most active voter is involved in the maximum number of brokerage roles. As he has access into the several communities of the wiki vote network being a broker, he has opportunity to nominate more people by casting vote in their favor. Moreover, presence of active voter in a number of communications taking place between different communities makes a sense of its high betweenness.

**Related Work**

This subsection provides insight into the available related studies in the perspective of Wikipedia elections. Researchers have noticed that Wikipedia has potential for election relevant activities. The role of Wikipedia has also been investigated in the context of offline as well as online elections by research community. Different statistical methods have been used previously to explore the Wikipedia role in the outcomes of offline elections (Smith and Gustafson 2017). For example, the importance of Wikipedia page views and polling data on US general Senate elections is explored in (Smith and Gustafson 2017). It is concluded that Wikipedia page views of a candidate can enhance the prediction of election outcomes successfully before election's day. It is also seen that creation of Wikipedia pages contribute to the electoral success of the contestants in US Congressional elections and UK parliamentary elections conducted in 2010 (Margolin et al. 2016). The effect of online information seeking at election time is also determined to find its correlation with the election results. It is shown that Wikipedia offers good insight into the election overall turnout and voting process for particular parties (Yasseri and Bright 2016). The authors argue that the number of page views on Wikipedia about elections can predict the possible turnout of voters. Moreover, Wikipedia page views, media news, and some details of political parties can influence the voting of parties participating in elections.

Different interesting observations have been made about the factors influencing the success of a candidate in Wikipedia in online settings. The available literature mostly focuses on the elections carried out for Wikipedia administrator selection. A group of administrators is selected by the community of contributors on Wikipedia. A person can request for adminship through a process named Request for Adminship (RfA). To initiate RfA process, a person must be registered as an editor on Wikipedia. Afterwards, s/he can nominate himself for adminship and also can be nominated by others through wiki public elections. People can vote for the persons of their own choice. After the expiry of the voting time period, results of the voting are reviewed and a final decision is concluded by bureaucrats (a special class of admins). The data about voters and nominations of wiki elections can be gathered by using Wikipedia page edit history.

The characteristics of candidates are identified and examined that can enhance their probability to be selected as administrator (Burke and Kraut 2008a). The factors on an individual level are focused to analyze the supporters of a candidate. Results show that strong edit history, edit summaries and wide-ranging experience of a candidate such as a user interaction contribute to the success of a candidate in the online elections. New admins of Wikipedia are discovered by modeling admin elections (Jankowski-Lorek et al. 2013). Wikipedia edit history is used in this study and classification of votes is done to select candidates for adminship. It is observed that in RfA process, voters cast vote for those persons who have relevant experience of editing articles on different topics. During the study of relative assessment used by voters in online elections, the relationship of a voter with the candidate is found important behind her/his voting decision (Leskovec et al. 2010c). Two factors named the *Total number of edits* and the *Number of total awards* received from community members are found important with respect to voters for the relative assessment of candidates.

Furthermore, it is found that triads in a social network can help to explore the voting behavior (Leskovec et al. 2010a). However, authors used the properties of a voter's social network as input features to predict the sign of a vote given by participants of the election. Two important predicators are found for the success of a candidate for being an admin (Burke and Kraut 2008b). First important factor, in this case, is the *Variety of experience* of that candidate in different areas such as article talk, Wikipedia talk, admin notice board, and other RfA's. The second one is the *Contribution* of candidates towards policy making for Wiki projects. The contributions of candidates are used to predict their success in online wiki elections (Kordzadeh and Kreider 2016). Total contribution, activity history, tenure, and the number of RfA attempts play an important role in user's selection as an admin. Also, voters are found positively contributing in voting for the contestants to whom they have already communicated (Lee et al. 2012).

## Conclusions

This study explores the importance of contacts of contestants in wiki elections. SNA is performed to examine personal and know-how contacts of a person. Besides, general and structural characteristics of the wiki-vote network are discussed. The results show that there is a great impact of personal contacts of an individual to be successful in online elections. Furthermore, know-how contacts of a person also influence the voting process of elections when he owns a significant position in information flow (whether in terms of having an intermediate position or agent role), and have relatively closer neighbors. Know-how contacts also influence the nominations for an administrator when a number of links, the quality of links, and input k-neighbors are considered. This study provides a roadmap to other researchers for data analysis of future wiki elections through the social network lens. There is a possibility of involving more persons in different brokerage roles and at different prominent positions in current wiki data.

In the future, we are interested in the collection of latest wiki-elections data and to find the change in the behavior of election participants. The factors behind the motivation or de-motivation of users in participation or the selection of administrator on Wikipedia network is the forthcoming intention. Another idea for future research is to conduct longitudinal studies to study the evolving behavior of contestants of online elections.

## Declarations

### List of abbreviations

SNA: social network analysis; RfA: request for adminship.

### Availability of data and material

The dataset supporting the conclusions of this article is available in the [SNAP STANFORD] repository, [http://snap.Stanford.edu/data/wiki-Vote.html].

### Competing interests

The authors declare that they have no competing interests.

**Funding**
The study was not funded.

**Authors' contributions**
**Yousra Asim**, written the complete draft of the paper, performed experiments and presented the results.
**Muaz A. Niazi**, given the initial idea of the paper about Wiki election and reviewed the entire paper.
**Basit Raza**, defined the structure of the paper and helped in scientific writing and reviewed the entire paper.
**Ahmad Kamran Malik**, provided the support in usage of tools for presenting different figures, technical writing and reviewed the entire paper.

**Acknowledgements**
This research work is supported by COMSATS Institute of Information Technology (CIIT) Islamabad, Pakistan.